\documentclass[12pt,preprint]{aastex}

\begin{document}
 
\title{Interferometric Observations of RS Ophiuchi and the Origin of the Near-IR Emission}
\author{B. F. Lane \altaffilmark{1},  J. L. Sokoloski\altaffilmark{2}, 
R. K. Barry\altaffilmark{3,4}, W. A. Traub\altaffilmark{5} ,A. Retter \altaffilmark{6},
M. W. Muterspaugh\altaffilmark{7}, R. R. Thompson\altaffilmark{8},
J. A. Eisner\altaffilmark{9}, E. Serabyn\altaffilmark{5}, B. Mennesson\altaffilmark{5}}

\altaffiltext{1}{Kavli Institute for Astrophysics and Space Research, MIT Department of Physics, 70 Vassar Street, Cambridge, MA 02139; blane@mit.edu}
\altaffiltext{2}{Columbia Astrophysics Laboratory; jeno@astro.columbia.edu}
\altaffiltext{3}{Johns-Hopkins University}
\altaffiltext{4}{NASA/GSFC, MC 667, NASA Goddard Space Flight Center, Greenbelt, MD 20771}
\altaffiltext{5}{JPL, M/S 301-451, 4800 Oak Grove Dr, Pasadena,CA 91109}
\altaffiltext{6}{Astronomy \& Astrophysics Dept., Penn State University, 525 Davey Lab, University Park, PA 16802-6305; retter@astro.psu.edu}
\altaffiltext{7}{Townes Fellow, Space Sciences Laboratory, University of California, Berkeley, CA 94720}
\altaffiltext{8}{Michelson Science Center, 100-22 California Institute of Technology, Pasadena, CA 91125;thompson@ipac.caltech.edu }
\altaffiltext{9}{Miller Fellow,  University ofCalifornia at Berkeley, 601 Campbell Hall, Berkeley, CA 94720}

\begin{abstract} 

We report observations of the recurrent nova RS Oph using
long-baseline near-IR interferometry. We are able to resolve emission
from the nova for several weeks after the February 2006 outburst. The
near-IR source initially expands to a size of $\sim 5$
milli-arcseconds.  However, beginning around day 10 the IR source appears 
to begin to shrink, reaching $\sim 2$ milli-arcseconds by day 100.  We combine
our measured angular diameters with previously available interferometric and 
photometric data to derive an emission measure for the source, and hence are able to
determine the mass-loss rate of the nova in the days following the
outburst.

\end{abstract}


\section{Introduction}

When a massive white dwarf (WD) accretes matter from a nearby star,
the material accumulates on the surface under increasing
temperatures and pressures. Once conditions for a thermonuclear
runaway reaction are met the system erupts as a nova \citep{pk05};
the luminosity increases by $\sim 9$--$19$ magnitudes over a few days
or weeks, and the system ejects mass in a strong, optically thick
wind. In most cases, the cycle repeats on
timescales that depend on the mass of the WD and the accretion rate --
``classical novae'' recur with periods of $\sim 10^4$ years, while
``recurrent novae'' erupt on decadal timescales.  It has been
argued \citep{hk99,hk01} that recurrent novae systems are the progenitors of
Type Ia supernovae, as models of the explosions indicate that the WD
should be very close to the Chandrasekhar limit, and that the WDs in
these systems are gaining more mass from the accretion than they are
losing from the outbursts \citep{kh04}.

RS Oph (HD 162214) is a recurrent nova which has undergone 6 recorded
outbursts since 1898, including eruptions in 1985 \citep{bode87} and
most recently in February 2006 \citep{IAUC8671}. In this system, the WD
has a red giant (RG) companion in a 460-day period circular orbit
\citep{fekel00}; mass loss from this companion provides fuel for the quiescent-state WD
accretion disk as well as a stellar wind that
interacts with the nova blast wave. Extensive observations of the two
most recent outbursts have been carried out at wavelengths ranging
from X-rays \citep{soko06,bode06} to radio \citep{IAUC8678}.

It should be noted that there has been some confusion as to the true
distance to the RS Oph system: most workers derive or assume a
distance of $\sim 1.6$ kpc, based primarily on radio observations
\citep{hjellming86,bode87,soko06}.  However, recent models \citep{hk00} of
this system indicated a distance of 0.6 kpc; \citet{monnier06}
noted that if the distance were indeed that small it might be possible
to resolve the two components of the system (the red giant --RG-- and the puffed-up
WD).  However, recently \citet{hk06b} revised their models and now
find a distance of 1.3--1.7 kpc. We will therefore take the system
distance to be 1.6 kpc.

 Recently, \cite{monnier06} used long-baseline infrared interferometry
to resolve the emission from the outburst. They find that the emission
is consistent with a source that is larger than the binary system.
Here we report on observations of the RS Oph system using the Palomar
Testbed Interferometer (PTI) on 15 nights following the 2006
outburst. We are able to resolve the emission on milli-arcsecond (mas)
scales and can place limits on the emission mechanism, including the
possibility that the emission is from the binary components. PTI was
 built by NASA/JPL as a testbed for developing
ground-based interferometry and is located on Palomar Mountain near
San Diego, CA. It combines starlight from two out of three available
40-cm apertures and measures the resulting interference fringes. The
angular resolution provided by this long-baseline (85-110 m), near
infrared ($1.2-2.2 \mu$m) interferometer is approximately 4 mas. A
complete description of the instrument can be found in
\cite{colavita99}.

\section{Observations \& Models}

We observed RS Oph on 15 nights between 24 March 2006 and 13 June 2006
(covering the period from 40 to 121 days after the outburst). Data
from 3 of these nights have been previously published in
\cite{monnier06} together with data from the IOTA \citep{iota} and
Keck interferometers \citep{colavita03}. We include all of these
observations in subsequent fits to help constrain models of this
source. In addition to the data obtained after the 2006 outburst, we
present data from an observation of the RS Oph system taken on 22 May
2003, i.e. before the outburst.

PTI operates using one interferometric baseline at a time. On the 2003
night and 8 of the 2006 nights RS Oph was observed using the
North-West baseline, the
remaining observations were done using the South-West baseline. 
The observing wavelength was 2.2 $\mu$m
(K~band; note that the IOTA data were obtained in the H band --
$\lambda = 1.6 \mu$m -- which may complicate inter-comparisons of the
data sets.)  Observations on the first 8 nights were done using a
fringe-tracking integration time of 20 ms, but as the source faded the
integration time had to be increased to 50 ms.  We made use of the
``coherent integration'' mode described in \cite{colavita99b}; the coherent
integration mode provides better performance at low flux levels. We also
performed model fits to the incoherently averaged data output and
verified that they gave consistent answers (but with larger
uncertainties.)

Each nightly observation with PTI consisted of one or more 130-second
integrations during which the normalized fringe visibility of the
science target was measured. These visibilities were calibrated by dividing 
them by the point-source response of the instrument, determined by
interleaving observations of
a calibration source: HD 161868 (A0V, $m_V=3.75$, $\theta = 0.69 \pm
0.1$ mas).  The angular diameter of the calibration source was
determined by fitting a black-body to archival broadband photometry.
For further details of the data-reduction process, see
\citet{colavita99b} and \citet{boden00}. The calibrated data, averaged 
by night, were listed in Table \ref{tab:data}; though given as averages,
the fits to the models were done using the individual data points. 

The normalized fringe visibility obtained from an interferometer is
related to the intensity distribution on the sky via the
van~Cittert-Zerneke theorem. For simple parametric models such as
uniform disks, gaussians or binary sources we use the procedure
outlined in \citet{eis03} to generate models that can be fit to the
observations. We perform least-squares fits of these models to the
observed fringe visibilities.

It is important to note that the IOTA data published by
\cite{monnier06} indicate a non-zero closure phase; this implies that
the emission is not point-symmetric. However, the amplitude of the
signal ($\sim 10$ degrees) also indicates that the degree of asymmetry
is not very large on scales resolved by that interferometer.  PTI,
being a single-baseline interferometer, is not able to measure closure
phases and therefore any model fits will have a 180-degree ambiguity
in position angle. We will consider intrinsically asymmetric models
(e.g. binaries), as well as symmetric ones -- in the latter case
recognizing that the model can only provide an approximation of the
true intensity distribution.

\section{Results}

\subsection{Post-Outburst Observations}

\begin{figure}[t]
\epsscale{1.0}
\plotone{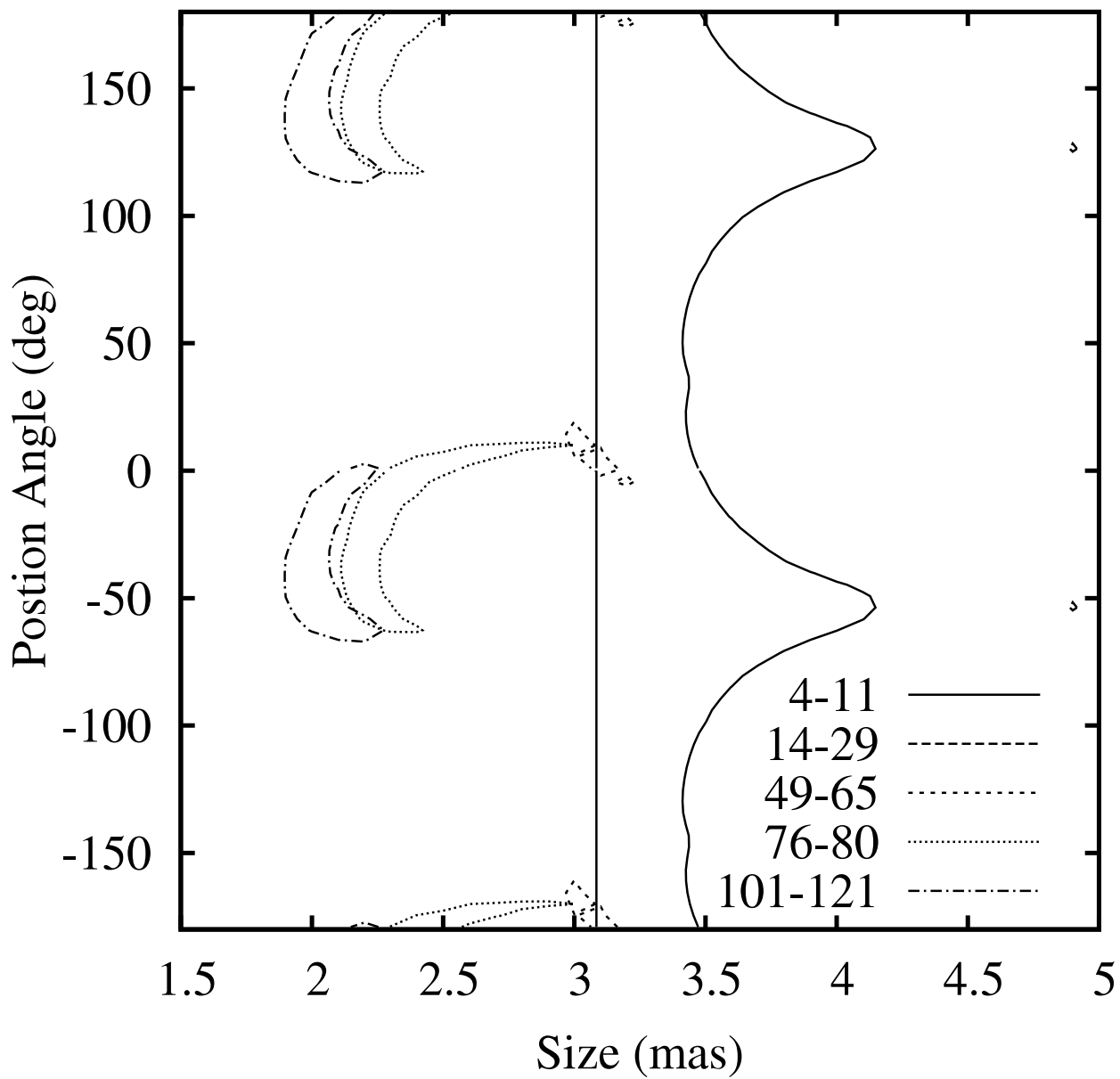}
\caption[]{\label{fig:chisurf} Plot of the $\chi^2$ surface of fits to
5 epochs of data on RS Oph, assuming an elliptical Gaussian intensity
distribution. The surface is a function of the size of the major axis
and position angle (degrees E of N) of the ellipse. The minor axis of
the fit was allowed to vary to minimize the $\chi^2$ and the best-fit
value is given in Table \ref{tab:fits}. The contours represent the
1-$\sigma$ uncertainty level, and the key indicates which contour
corresponds to which epoch.}
\end{figure}

We find that the data can be reasonably well fit by a simple
elliptical Gaussian model (Table \ref{tab:fits}); the fit is further improved when the
parameters are allowed to vary slowly in time.  Given that all the PTI
data from any given night are limited to a single baseline, it is not
possible to constrain complex models such as elliptical Gaussians on a
night-by-night basis.  We therefore group the available data into sets
of several nights, comprising observations using multiple
baselines, and fit for size, aspect ratio and position angle of the
long axis. Results are shown in Table \ref{tab:fits} and Figure
\ref{fig:chisurf}.  The $\chi^2$ surface indicates that the emission
morphology first expanded in size, with a position angle of $\sim 120$
degrees; subsequently the orientation changes to $\sim 0$ and the
source begins contracting. The apparent angular diameter of the major
axis reaches 5 mas at maximum extent, but contracts to 2 mas by day
100. It should be emphasized that that these observations trace the 
emission and not the material itself; as will be discussed in 
Section \ref{discussion} the apparent contraction does 
not necessarily require infall of material.

\begin{figure}[t]
\epsscale{1.0}
\plotone{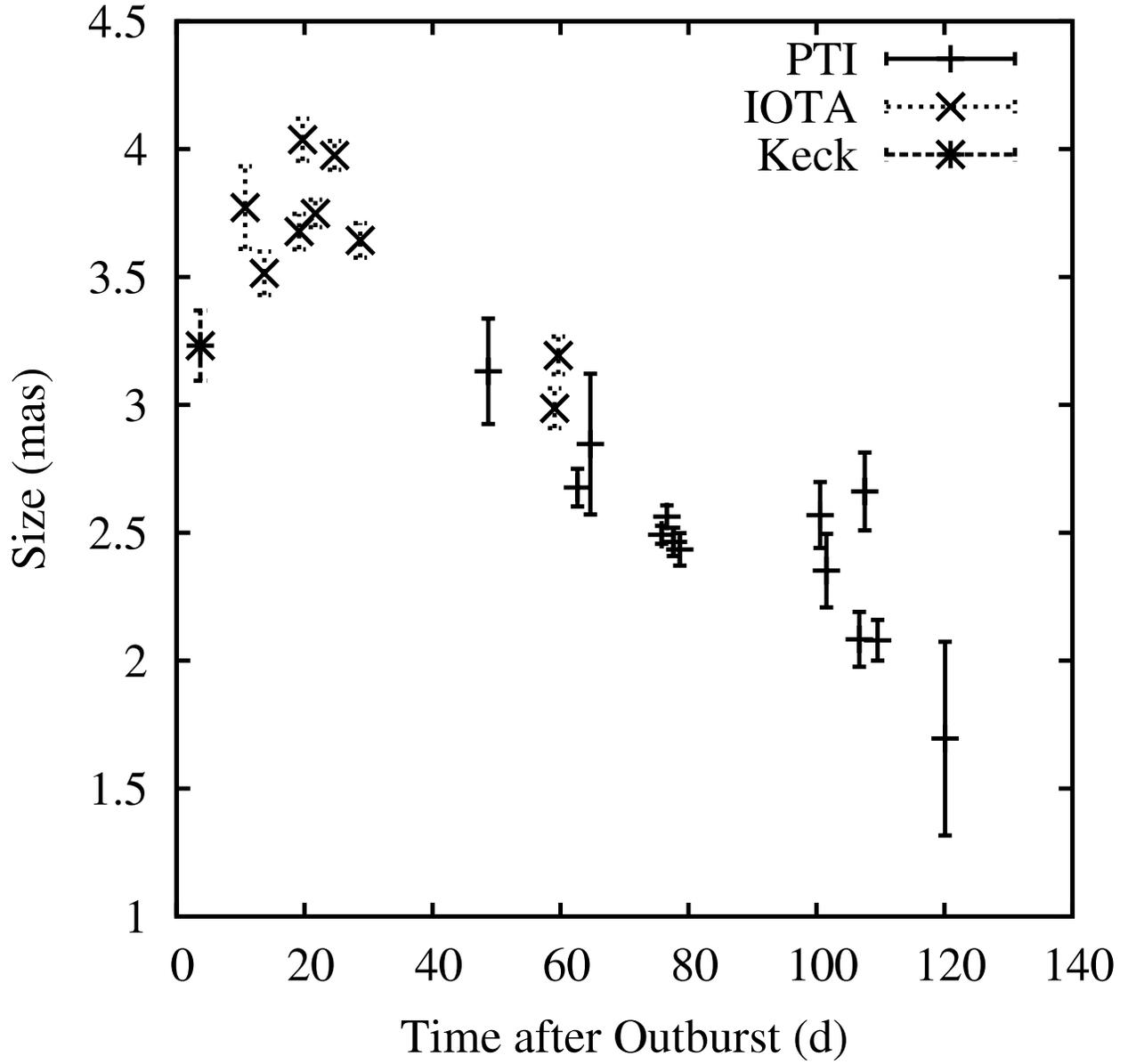}
\caption[]{\label{fig:diam} Best-fit night-by-night angular diameter of the 
major-axis of Gaussian emission model, with the 
orientation and position angle constrained to the overall best-fit 
values explained in the text.}
\end{figure}

We explore the changing angular size of the emission by constraining
the aspect ratio to the best overall value (Major/Minor = 1.4) and
position angle (PA=0) of the models and fitting for the size of the
corresponding elliptical gaussian on a night-by-night basis.  The
results are shown in Figure \ref{fig:diam}.

It is also possible to fit the data from each epoch using a
two-component model (indeed there is a degeneracy between a partially
resolved two component model and an elongated disk).
\citet{monnier06} find that their early-epoch RS Oph data are
consistent with a pair of components separated by $\sim 3$ mas.
However, if one is in fact seeing a "reborn" binary then the
two-component model should obey a Keplerian constraint. Hence we fit a
Keplerian binary model constrained to the orbital period (P=460 days),
eccentricity (0) and epoch ($T_0=2450154.1$) determined by radial velocity
observation \citep{fekel00}. We search for the best-fit semi-major
axis, inclination, longitude of ascending node, and component
intensity ratio. We find no acceptable fit with a semi-major axis
smaller than 2.3 mas, corresponding to $\sim 3.7$ AU. Thus 
we conclude that we cannot be seeing the binary itself.  

\subsection{Pre-Outburst Observations}

The observation in 2003 is comparatively limited in that it comprises
only two 130-second integrations on one night, using a single
baseline. The resulting $uv$-plane coverage is so limited as to
preclude meaningful fits to anything but the simplest uniform disk
model. However, the data are consistent with a single essentially
unresolved source with an angular diameter of $\theta_{UD} = 0.7 \pm
0.5$ milli-arcseconds (the large error bar comes from the fact that
$\theta_{UD} \ll \lambda/B$).

\section{Discussion}
\label{discussion}
The interferometric observations from IOTA and PTI directly constrain
certain aspects of the near-IR emission. First, the shortest
baseline data (from IOTA, baseline length $\sim 10.5$ meters) taken on
days 4--11 indicate a fringe contrast near unity ($V\sim 0.99$; see
Fig 1 in Monnier et al. 2006). This places an upper limit on the
amount of emission that can be coming from a source
larger than the interferometric field of view (
$\lambda^2/B\Delta\lambda \sim 0.08$ arcsec for IOTA).  That upper
limit is $\sim 0.7$\%.  Second, the long-baseline observations indicate that the
source is in fact resolved; for the longest baselines $V^2 \sim 0.4$,
indicating a size on the order of $\sim \lambda/B \sim 3$ mas.  For
reference, the blast wave from the nova expanded at $\sim 3,500~ {\rm
km s^{-1}}$ \citep{soko06}, implying an angular diameter of $5$ mas by
day 4, and an expansion rate of $\sim 1.2$ mas/day. 

Let us now consider the source of the observed extended near-IR
emission.  Given the measured angular size scale of $\sim 3$ mas, the
physical scale is $\sim 2$ AU, slightly larger than the expected
orbital separation of 1 AU \citep{fekel00}. One possible source of extended near-IR
emission is blackbody radiation from optically thin dust surrounding
the RS Oph system.  However,
recent spectroscopic observations in the near- and mid-IR have not
found any evidence for dust emission \citep{IAUC8710}. In addition, we
point out that if there were dust, it is hard to explain the apparent
contraction of the emission source; if the dust is evaporating it
would likely evaporate from the inner regions first - yielding an
''expanding ring'' morphology.

Another possible source of the extended near-IR emission
is the post-outburst wind emanating from the nova. It is expected that
after the initial explosion shell material on the WD will undergo
hydrogen burning for weeks to months; the energy thus liberated will
cause most of the shell material to leave the WD in a strong stellar
wind \citep{kh94}. Radio observations of the ejecta indicate an
asymmetric morphology with a jet-like feature at a position angle near
90 deg; this is reasonably consistent with the inital elongation seen
in our data (P.A. $\sim 120$) that may represent the base of that jet
\citep{ob06}. In this scenario the initial elongation would represent 
faster-moving material leaving along the polar axis of the system; 
subsequent, slower material would be concentrated along the system 
orbital plane.  It is important to remember that the mere fact that the emission 
region appears to shrink does not necessarily imply that the 
material is falling in; what we are seeing is the effective photospheric
diameter, which is a function of the density of the material. If the 
mass-loss rate from the WD drops then the apparent photospheric 
radius would be expected to shrink even as the material continues 
flowing outward. 

A post-nova wind would be expected to radiate free-free, bound-free
and line emission \citep{rybicki,pearson05}, and proper treatment would require a full
photo-ionization model. However, a simplified treatment
using only the free-free emission should be adequate for initial mass
estimates; in particular it should produce an upper limit. In this context we note that PTI is
equipped with a low-resolution spectrometer which disperses light in
the K band across 5 channels, providing both photometry and fringe
visibilities in each channel. A close inspection of the spectrally resolved data
reveals no significant change in fringe visibility, nor any significant excess 
emission, in the channels that would contain known strong emission 
lines (e.g. Br$\gamma$, 2.166 $\mu$m); the level of photometric uncertainty 
is $\sim 5$\%.  This does not exclude line emission, but indicates that the total flux seen by PTI
is dominated by continuum emission. This is consistent with previously
measured equivalent widths in the Br$\gamma$ line \citep{callus86}.
It is also consistent with H and K-band IR spectra by \citet{evans06} which 
indicate that the line/continuum ratio is in the range 10--30 \% in the period 
from 11 to 55 days after the eruption (with the ratio decreasing over time).

We use previously published IR photometry \citep{evans88} and
reddening estimates of $E(B-V)=0.73 $ from \citet{IAUC4067}, obtained during the 1985
outburst to constrain the emission model, justified by the fact that 
we find that the 1985 light curve is consistent with the 2006 K-band 
light curve as measured by PTI. In addition to the postulated wind emission
 source, there are other sources of IR radiation: the M-giant companion and 
 possibly the WD accretion disk. However, we expect that the accretion disk will in
fact be disrupted by the nova and not reform for $\sim 1$ year, based
on the fact that the post-outburst light-curve drops below the mean
long-term intensity, and the cessation of rapid flickering that is
usually associated with accretion disks \citep{evans88,z06}. We are
therefore only left with the M-giant companion; we attribute all of
the near-IR emission at the minimum intensity point in the light-curve
to it. 

\begin{figure}[t]
\epsscale{1.0}
\plotone{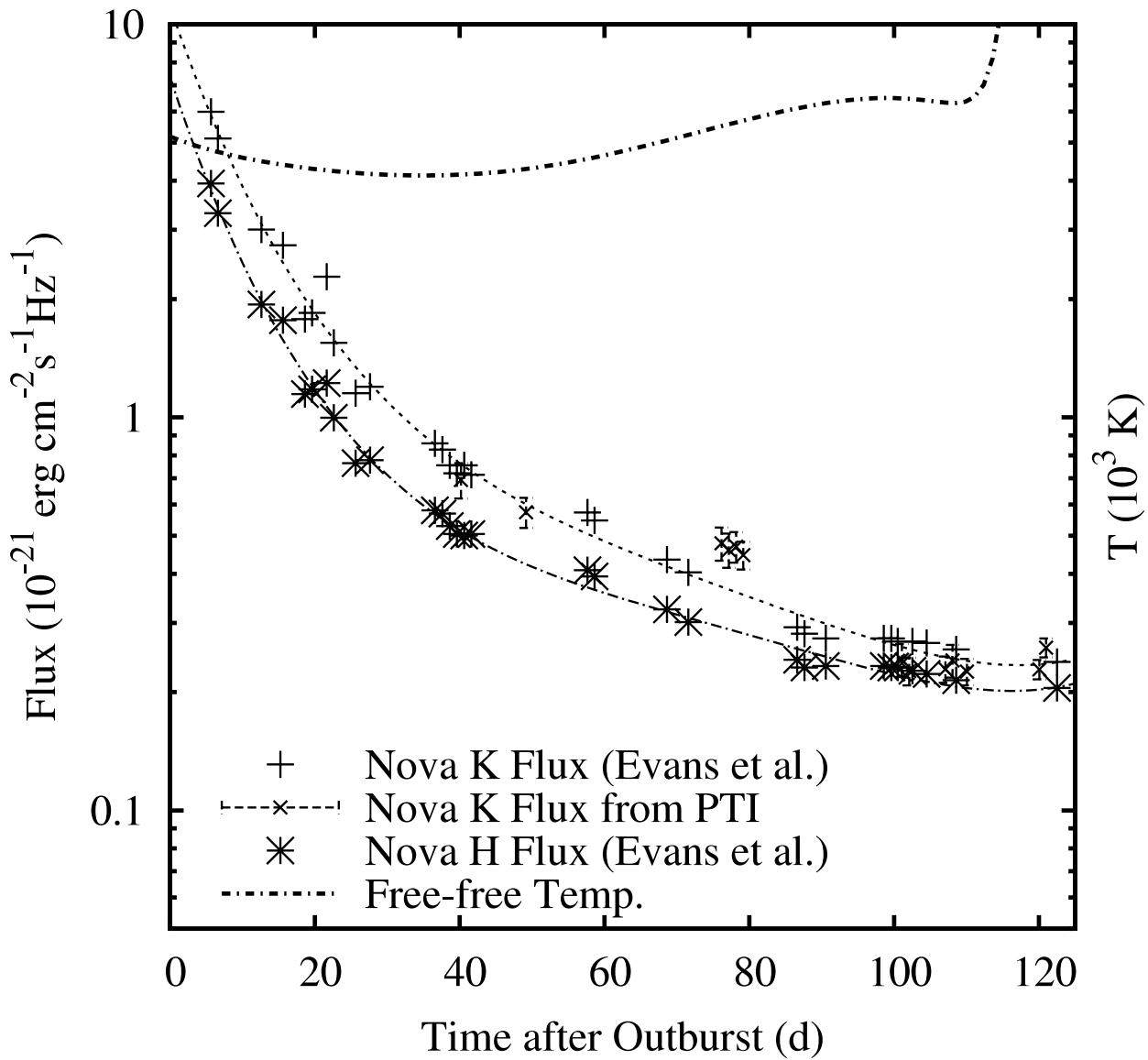}
\caption[]{\label{fig:phot} The near-IR fluxes from the nova source as
measured by PTI and from \cite{evans88} together with the derived
color temperature assuming free-free emission. Low-order polynomials
have been fit to the photometry for use in interpolation.}
\end{figure}

After subtracting the flux from the companion, we
derive the H and K-band intensities of the nova source. The H/K flux
ratio thus derived can be used to determine the emission temperature
for the thermal Bremsstrahlung process ($f(H)/f(K) \simeq \exp(h(\nu_H - \nu_K)/kT)$
neglecting variations in the Gaunt factors);
this yields a temperature in the range of $\sim 10^4$ K (Fig
\ref{fig:phot}).  Note that for an apparent angular diameter of $\sim
3$ mas, in order to match the angular size, brightness and color
temperature the source must be optically thin \citep{ney80}.

\begin{figure}[t]
\epsscale{1.0}
\plotone{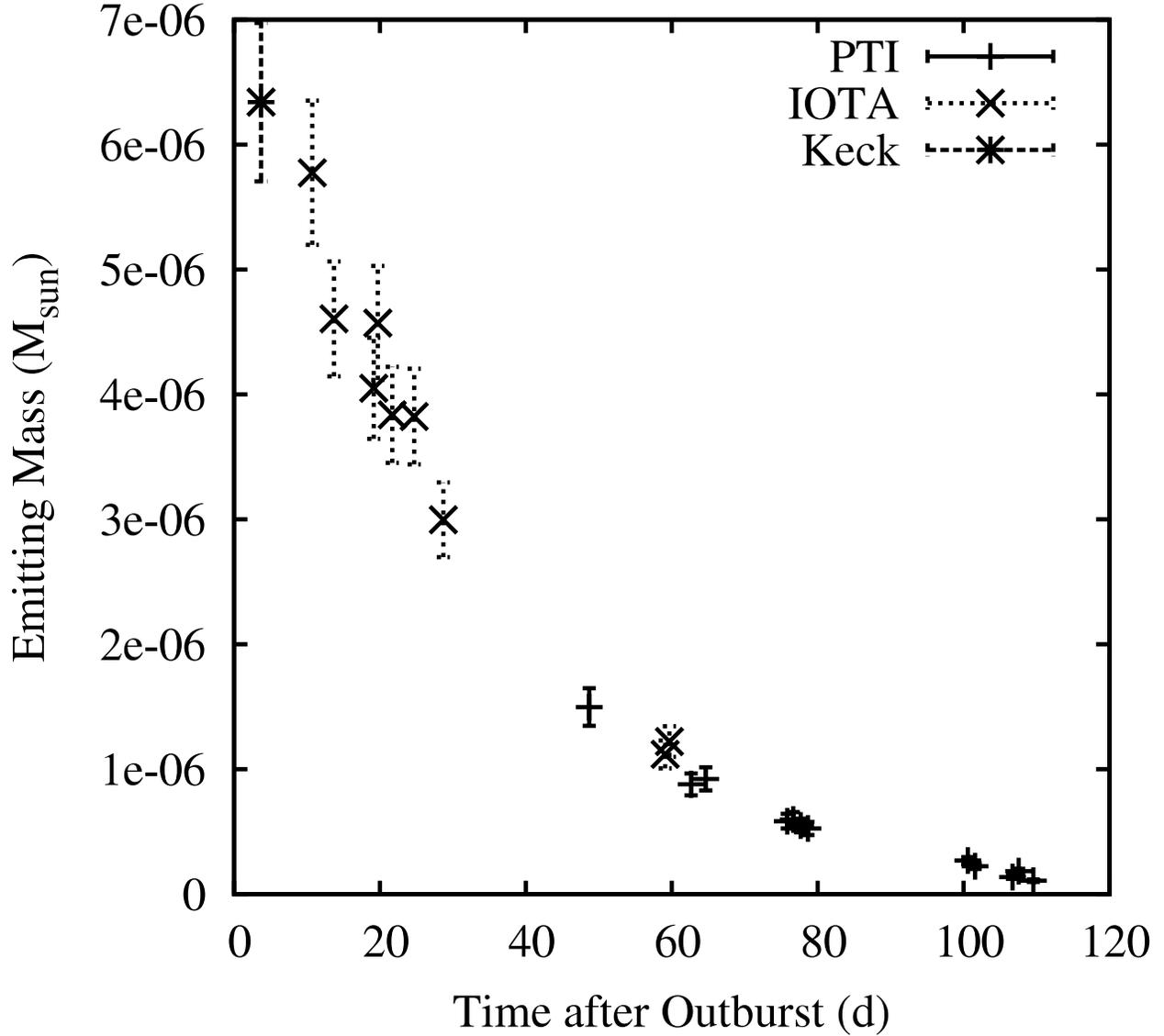}
\caption[]{\label{fig:mass} The mass of the near-IR emission source as
a function of time, assuming thermal free-free emission. The emission
measure was determined from the 1985 light curve, while the angular
diameter measurements allow us to solve for $n_e^2$ and total mass
separately, assuming a uniform density distribution.}
\end{figure}

For thermal Bremsstrahlung the emitted intensity is proportional to
$\int n_e^2 dV$, i.e. the emission measure. Using the
interferometrically determined sizes we can find the total volume $V$
and hence infer the electron density and total emitting mass as a
function of time, assuming that the electron density is uniform
throughout the emitting volume. Results are shown in Figure
\ref{fig:mass}. In determining the volume, we assume the emitting
region to be an ellipsoid of revolution with dimensions equal to the
major and minor axes of the best-fit Gaussian model.

We find that the required emitting mass is in the range of 1--6
$\times 10^{-6} M_{\odot}$. If this material is being ejected from
the nova at high velocity then the total amount of mass ejected 
in the months after the eruption becomes
considerable: given the size of the emitting region, and
a nominal wind velocity of $\sim 1000 {\rm km s^{-1}}$, material will
cross the emitting region in a few days. We find the approximate total
mass ejected from
\begin{equation}
M_{ej} \simeq \frac{T_{ej} m v_{exp}}{R}
\end{equation}
where $T_{ej}$ is the duration of mass ejection \citep[$\sim 60$ days]{hk06b}, $m$
is the amount of mass in the emitting region at any given time, $v_{exp}$ is the
expansion velocity and $R$ is the size of the emitting region. Hence
the total amount of mass ejected post-outburst is $\sim 6 \times
10^{-5}  (T_{ej}/60{\rm d})(v_{exp}/1000 {\rm km.s}^{-1}) M_{\odot}$. Such a large ejected mass is consistent with the shell
mass determined by \citet{boh89}, but is larger that the predicted
value from models ($10^{-6}$--$10^{-7} M_{\odot}$; Yaron et al. 2005.),
and larger than estimates based on X-ray observations from the 
1985 outburst \citep{ob92}.  We note that the mass estimate of a few $ \times 10^{-7} M_{\odot}$ by 
\citet{soko06} is based on the X-ray properties of the shock in the period immediately after 
the outburst and would  not be sensitive to mass ejected many days after 
optical maximum, as seen here. Nevertheless, this is comparatively large amount of mass, roughly 
equal to the amount of mass lost from the red giant in the inter-outburst period;
this would cast doubt on the notion that the nova is a net gainer of mass. 

One can however propose other mechanisms to explain the presence of a 
few $\times 10^{-6} M_{\odot}$ of emitting material around the nova binary system,
if this material is considered to be quasi-stationary.  Two sources 
of material are readily available in the system: the accretion disk and
the red giant. The first scenario would be that the material we are seeing comes from 
the accretion disk that surrounds the WD (the disk is thought to be disrupted
by the nova); if this material is heated by the passing shock wave and
enough momentum is imparted to move it outside of the binary, it might
be possible to match the observed morphology, i.e the elongation with
P.A. $\sim 0$ would then be aligned with the orbital plane (inferred
by \citet{ob06} to be perpendicular to the jet direction.) If we are
indeed seeing gas emission by material from a disrupted disk, then these measurements
represent a measurement of the mass of that disk. If we assume that
the material is quasi-stationary then the inferred disk mass is 1--6
$\times 10^{-6}  M_{\odot} $.

The second possibility is that the blast wave from the nova impacts the 
red giant with sufficient momentum to strip off a portion of the companion 
atmosphere. Such scenarios have been studied in the context of supernovae 
\citep{mar00,w76}, and it has been shown that considerable mass stripping
will occur. However, the greatly reduced energy available in novae 
limits the magnitude of this effect. Assuming a blast wave velocity of 
$3000~ {\rm km.s}^{-1}$ and mass of $\sim 5 \times 10^{-7} M_{\odot}$ \citep{soko06},
and a RG radius of $\sim 30 R_{\odot}$, mass of $\sim 0.7 M_{\odot}$
and the orbital separation $\sim 320 R_{\odot}$ \citep{hk06}, and using
Eqns 8 \& 9 from \citet{w76}, we find that their parameter $\Psi \sim 5 \times 10^{-8}$ ($\Psi$ is
essentially the ratio of the momentum of the portion of the blast wave that impacts 
the companion to the momentum required to move the companion at its own
escape velocity). \citet{w76} find that the fraction of mass stripped from the 
companion is $\propto \Psi^3$; hence we conclude that the amount of mass
stripped from the RG is essentially negligible. 

\section{Conclusion}

We have used long-baseline interferometry to resolve the emission from
the recurrent nova RS Oph. We find that the near-IR emission is on the
scale of a few mas, corresponding to $\sim 2$ AU at the canonical 1.6
kpc distance; this is factor of $\sim 50$ smaller that the size of the
radio emission source at this time, but is also larger than the size of the 
WD-RG binary system. The emission is also not
spherically symmetric. Initially the source appeared to expand and was elongated with a
major-axis position angle of $\sim 120$ deg; subsequently the source appears to contract 
and the orientation angle changes to $\sim 0$.  We explore possible sources of
this emission, and attribute it to plasma emission emanating from the
strong wind leaving the WD. We derive an upper limit to the mass in the emitting region
to be a few times $10^{-6} M_{\odot}$ and use it to infer the total
amount of mass lost by the WD during the days immediately following
the outburst to be $\sim 6 \times 10^{-5} (T_{ej}/60{\rm d})(v_{exp}/1000 {\rm km.s}^{-1}) M_{\odot} $.

\acknowledgements We are grateful to A. Evans for providing us 
with computer-readable spectra. We wish to acknowledge the remarkable
observational efforts of K. Rykoski. Observations with PTI are made
possible through the efforts of the PTI Collaboration, which we
gratefully acknowledge. This research has made use of services from
the Michelson Science Center, California Institute of Technology,
http://msc.caltech.edu.  Part of the work described in this paper was
performed at the Jet Propulsion Laboratory under contract with the
National Aeronautics and Space Administration. This research has made
use of the Simbad database, operated at CDS, Strasbourg, France, and
of data products from the Two Micron All Sky Survey, which is a joint
project of the University of Massachusetts and the Infrared Processing
and Analysis Center/California Institute of Technology, funded by
NASA and the NSF. BFL acknowledges support from a Pappalardo
Fellowship in Physics, and JLS is supported by
an NSF Astronomy and Astrophysics Postdoctoral Fellowship under award AST-0302055.

\clearpage
\begin{deluxetable}{l c c c c c c}
\label{tab:data}
\tablecaption{Observed Fringe Visibilities\label{tab:data}}
\tablewidth{0pt}
\tablehead{
Date& \colhead{Julian Date (MJD)} & \colhead{U\tablenotemark{a}}&
\colhead{V\tablenotemark{a}} & \colhead{No. Pts.\tablenotemark{b}} & \colhead{$V^2$\tablenotemark{c}} &\colhead{$\sigma_{V^2}$}
}
\startdata 
5/22/2003 & 52781.384 & -83.2191 & -20.7015 &  2 & 0.970 & 0.242 \\ 
3/24/2006 & 53818.510 & -83.1262 & -22.8196 &  1 & 0.126 & 0.050 \\ 
4/2/2006   & 53827.510 & -83.2183 & -21.2903 &  3 & 0.337 & 0.077 \\ 
4/16/2006 & 53841.491 & -81.0884 & -20.1695 &  9 & 0.411 & 0.031 \\ 
4/18/2006 & 53843.491 & -81.5573 & -19.8387 &  8 & 0.300 & 0.050 \\ 
4/29/2006 & 53854.464 & -80.9573 & -19.6682 &  8 & 0.459 & 0.029 \\ 
4/30/2006 & 53855.456 & -81.1711 & -19.9687 & 10 & 0.448 & 0.029 \\ 
5/1/2006  & 53856.481 & -49.2736 &  58.0854 &  9 & 0.366 & 0.058 \\ 
5/2/2006  & 53857.478 & -49.3282 &  58.0682 &  9 & 0.389 & 0.026 \\ 
5/24/2006  & 53879.403 & -80.2290 & -19.2295 &  6 & 0.493 & 0.144 \\ 
5/25/2006 & 53880.407 & -47.8541 &  57.7800 &  7 & 0.424 & 0.072 \\ 
5/30/2006 & 53885.390 & -46.9012 &  57.6591 &  6 & 0.556 & 0.074 \\ 
5/31/2006 & 53886.380 & -80.9950 & -19.4490 &  5 & 0.458 & 0.051 \\ 
6/2/2006 & 53888.377 & -45.7106 &  57.4739 &  4 & 0.485 & 0.094 \\ 
6/12/2006 & 53898.374 & -51.3773 &  58.3395 &  2 & 0.735 & 0.454 \\ 
6/13/2006 & 53899.328 & -82.8230 & -20.4248 &  3 & 0.892 & 0.078 \\ 
\enddata
\tablenotetext{a}{Projected baseline coordinates in meters.}
\tablenotetext{b}{Number of 130-second integrations averaged 
together for the night.}
\tablenotetext{c}{Normalized fringe visibility, squared.}
\tablecomments{Table of observed fringe visibilities,
averaged on a nightly basis.}
\end{deluxetable} 

\clearpage
\begin{deluxetable}{l c c c c c c}
\tablecaption{Model Fits\label{tab:fits}}
\tablewidth{0pt}
\tablehead{
\colhead{Model} & \colhead{All PTI }& \colhead{Epoch 1} & \colhead{Epoch 2} & \colhead{Epoch 3} &\colhead{Epoch 4} & \colhead{Epoch 5}\\
\colhead{Parameters }  & \colhead{data}      & \colhead{(day 4-11)} & \colhead{(day 14-29)} & \colhead{(day 49-65)} & \colhead{(day 76-80)}   & \colhead{(day 101-121)} 
}
\startdata 
Ellip. Gauss.&           &            &             &             &               &   \\
Major Axis   &$2.6\pm0.9$&$3.2\pm0.1$ &$5.88\pm0.4$  & $2.7\pm0.2$ & $2.1\pm0.23$  & $2.0\pm0.8$  \\ 
Minor Axis   &$1.7\pm0.7$&$1.9\pm1.3$ &$3.4\pm0.1$  & $1.5\pm0.02$& $2.0\pm0.1$   & $1.0\pm0.4$  \\ 
P.A.         &$11 \pm 14$&$29\pm 14$  &$127\pm2$  & $10\pm42$   & $149\pm445$   & $-25\pm343$  \\ 
$\chi^2_r$   & 1.9       & 1.4        & 2.4         &  4.4        & 1.1           & 1.5          \\
No. Pts.     & 90        & 22         & 108         &  52         & 36            & 33           
\enddata

\end{deluxetable} 

\clearpage

\clearpage

\clearpage

\clearpage

\end{document}